\documentclass[12pt, draftclsnofoot,onecolumn]{IEEEtran}
\usepackage{dsfont}
\usepackage{amssymb}
\usepackage{subfig}
\usepackage{graphicx, amssymb, amsmath}
\usepackage{extarrows}
\usepackage{color}
\emergencystretch=\maxdimen
\IEEEoverridecommandlockouts
\begin{document}

\title{Self-Sustaining Caching Stations: Towards Cost-Effective 5G-Enabled Vehicular Networks}

\author{Shan~Zhang,~\IEEEmembership{Member,~IEEE,}
        Ning~Zhang,~\IEEEmembership{Member,~IEEE,}
        Xiaojie~Fang,~\IEEEmembership{Student~Member,~IEEE,}
        Peng~Yang,~\IEEEmembership{Student~Member,~IEEE,}
        and~Xuemin~(Sherman)~Shen,~\IEEEmembership{Fellow,~IEEE}
\thanks{Shan~Zhang and Xuemin~(Sherman)~Shen are with the Department of Electrical and Computer Engineering, University of Waterloo, 200 University Avenue West, Waterloo, Ontario, Canada, N2L 3G1 (Email:\{s327zhan, sshen\}@uwaterloo.ca).}
\thanks{Ning~Zhang is with the Department of Computing Science, Texas A\&M University-Corpus Christi, 6300 Ocean Dr., Corpus Christi, Texas, USA, 78412 (Email: zhangningbupt@gmail.com).}
\thanks{X. Fang is with the School of Electronics and Information Technology, Harbin Institute of Technology, Harbin, China (email: fangxiaojie@hit.edu.cn).}
\thanks{Peng Yang is with the School of Electronic Information and Communications, Huazhong University of Science and Technology, Wuhan, China (email: yangpeng@hust.edu.cn).}
\thanks{This work is sponsored in part by the Nature Science Foundation of China No. 91638204 and the Natural Sciences and Engineering Research Council of Canada.}
}

\maketitle

\begin{abstract}
	
	In this article, we investigate the cost-effective 5G-enabled vehicular networks to support emerging vehicular applications, such as autonomous driving, in-car infotainment and location-based road services.  
	To this end, self-sustaining caching stations (SCSs) are introduced to liberate on-road base stations from the constraints of power lines and wired backhauls.
	Specifically, the cache-enabled SCSs are powered by renewable energy and connected to core networks through wireless backhauls, which can realize ``drop-and-play'' deployment, green operation, and low-latency services.
	With SCSs integrated, a 5G-enabled heterogeneous vehicular networking architecture is further proposed, where SCSs are deployed along roadside for traffic offloading while conventional macro base stations (MBSs) provide ubiquitous coverage to vehicles.
	In addition, a hierarchical network management framework is designed to deal with high dynamics in vehicular traffic and renewable energy, where content caching, energy management and traffic steering are jointly investigated to optimize the service capability of SCSs with balanced power demand and supply in different time scales.
	Case studies are provided to illustrate SCS deployment and operation designs, and some open research issues are also discussed.

\end{abstract}


\section{Introduction}
Vehicular communication networks hold the promise to improve transportation efficiency and road safety, by enabling vehicles to share information and coordinate with each other.
Several potential vehicular networking solutions have been proposed, such as IEEE 802.11p standard and cellular-based techniques \cite{Abboud16_VANET_interworking}.
Compared with other candidates, cellular-based vehicular networking can benefit from the existing cellular network infrastructures to provide ubiquitous coverage and better quality of service (QoS) \cite{Sun16_LTEV_mag}.
In fact, 80\% of on-road wireless traffic is served by cellular networks \cite{Malandrino16_OnRoadFog_data}.
Therefore, cellular-based vehicular networking has drawn extensive attention from both academia and industry.
Specifically, the 3rd Generation Partnership Project (3GPP) is currently specifying LTE enhancements to support both vehicle-to-vehicle (V2V) and vehicle-to-infrastructure (V2I) communications, by integrating cellular and device-to-device interfaces \cite{Seo16_LTEV_standard_mag}.
The corresponding specification work will be finalized as a part of Release 14 in 2017, which can provide a full set of technological enabler from air interface to protocols.
In addition, extensive LTE-vehicular trail testing are now on-going in different places, like Germany and China.
Particularly, LTE-based vehicular networking has demonstrated its advantages in achieving significant message coverage gain, compared with IEEE 802.11 technologies in both high speed highway and congested urban scenarios \cite{Ericsson_LTEV}.

Despite the favorable advantages, cellular networks still face tremendous challenges to meet the needs of future vehicular communications, and the most pressing one is network capacity enhancement.
Currently, the on-road wireless traffic accounts for 11\% of the cellular traffic \cite{Malandrino16_OnRoadFog_data}, which expects to dramatically increase due to the proliferation of connected vehicles and emerging applications such as autonomous driving, in-car infotainment, augment reality, and location-based road services.
Deploying on-road base stations is the most effective way to increase vehicular network capacity.
However, conventional base stations requires power lines and wired backhaul connections, making on-road deployment greatly challenging and costly.
Furthermore, densification of on-road base stations may also lead to huge energy consumption and bring heavy burdens to backhauls, which can increase operational cost and degrade service performance.

In this article, we first introduce a new type of 5G-enabled on-road base station, namely self-sustaining caching stations (SCSs), to enhance vehicular network capacity in a cost-effective way.
Specifically, the SCSs have three features: (1) powered by renewable energy instead of power grid, (2) connected to the core network via millimeter wave (mmWave) backhauls, and (3) cache-enabled for efficient content delivery.
By leveraging these 5G technologies, SCSs can be deployed flexibly in a ``drop-and-play'' manner without wired connections, enable green network operation without additional on-grid energy consumption, and improve delay performance by relieving backhaul pressures.
Then, we propose a cost-effective heterogeneous vehicular network architecture, where SCSs are deployed along roadside to enhance network capacity while conventional macro base stations provide ubiquitous coverage and control signaling.

To harness the potential benefits of proposed network architecture, we further design a hierarchical management framework to deal with challenges such as intermittent renewable energy supply and highly dynamic traffic demand.
Particularly, cached contents are updated to maintain content hit rate considering vehicular mobility, while energy management and traffic steering are performed to balance power demand and supply in both large and small time scales.
In addition, case studies are provided to illustrate the implementation of proposed architecture in details, including cache size optimization and sustainable traffic-energy management.

The remainder of this article is organized as follows. In Section~\ref{sec_architecture}, the basics of SCSs is introduced, based on which a heterogeneous vehicular network architecture is proposed. Then, a hierarchical network management framework is designed in Section~\ref{sec_management}, and case studies are provided in Section~\ref{sec_CaseStudy}. Finally, Section~\ref{sec_research_topic} discusses future research topics, followed by the conclusions in Section~\ref{sec_conclusions}.

\section{Vehicular Network Architecture with SCSs}
    \label{sec_architecture}
	\subsection{{{Cellular-Based Vehicular Networks}}}
	
	With existing infrastructures and the state-of-the-art technical solutions, cellular-based vehicular networks hold the promise to provide ubiquitous coverage and support comprehensive QoS requirements in different scenarios. 
	For example, the hidden terminal problems of the 802.11p standard can be totally avoided \cite{Omar13_vemac}.
	Besides, low latency and high reliability can be guaranteed even in dense traffic scenarios, with effective congestion control and resource management schemes.
	
	In spite of the aforementioned advantages, cellular-based vehicular networks still face significant challenges.
	With the rapid development of information and communication technologies, massive advanced on-road technologies and applications are emerging, such as autonomous driving, argument reality, infotainment services, and other location-based road services.
	As these data-hungry applications will bring the surge wireless traffic, improving vehicular network capacity has become an urgent issue.
	To this end, on-road base stations need to be deployed.
	However, conventional base stations are connected in a wired manner due to the requirement of on-gird power supply and backhaul transmission, which can cause the following problems. 
	{{Firstly, conventional base stations rely on power lines and wired backhaul (e.g., optical fiber) to function, resulting in inflexible deployment especially in areas with undeveloped power lines or fiber connections (such as highways and rural areas). 
			Secondly, the huge energy consumption can cause high operational expenditure as well as environmental concerns.
			Furthermore, with popularity in multimedia and localized services on the wheel, conventional base stations only offering connectivity might fail to provide satisfying QoS, due to the time-consuming file fetching from remote servers.}} 
	
	\subsection{Self-Sustaining Caching Stations (SCS)}
	
	Considering the characteristics of vehicular networking and emerging on-road applications, we leverage promising 5G technologies and propose to deploy SCSs in addition to existing cellular networks to enhance vehicular network capacity in a cost-effective way.
	Specifically, SCSs are equipped with energy harvesting techniques and content caching units, which are connected to the core network through mmWave wireless backhuals. 
	
	Equipped with solar panels or wind turbines, SCSs can harvest renewable energy to operate in a self-sustaining manner without the support of power grid\footnote{The typical solar panel with 15\% conversion efficiency can harvest 100W energy only by a 82 cm $\times$ 82 cm solar panel under rated sunlight radiation, which is sufficient to power a micro (/pico) base station with power demand of 80 W (/8 W). \cite{EH_net_cost_2013}}.
	Exploiting renewable energy as a supplementary or alternative power sources is an inevitable trend in the 5G era and beyond, as wireless network energy efficiency expects to be improved by 1000 times \cite{5G_Overview_JSAC14_JAndrews}.
	In addition, renewable energy harvesting can liberate network deployment from power lines.
	Wireless backhaul can be supported by the mmWave wireless communication technologies.
	With large bandwidth unlicensed, mmWave bands can realize broadband wireless communication based on the massive multiple input multiple output (MIMO) and beamforming technologies \cite{Qiao_proactive_mmWave}.
	Therefore, SCSs, which combine both energy harvesting and mmWave backhaul techniques, can be deployed in a ``drop-and-play'' manner with no wired constraints.
	
	Content caching empowers SCSs to store popular contents at the edge of networks, and thus reduce duplicate transmission from remote servers.
	As a matter of fact, the main on-road mobile applications is now generated by video streaming and map services, which are responsible for 80\% and 15\% of total traffic, respectively.
	The popularity of video contents has been found to follow power-law distribution.
	Accordingly, caching popular video contents in SCSs can effectively offload traffic from existing cellular systems.
	Moreover, the emerging on-road applications are expected to be location-based with concentrated request, which further make a strong case for content caching.
	In addition to capacity enhance, caching can also reduce transmission latency and relieve backhaul burdens, with contents stored closer to end users.
	Furthermore, caching schemes can be devised with respect to specific objectives, such as mobility-aware caching.
	Specifically, the content can be pre-fetched and stored in the next cells before the vehicles conduct handover, to realize smoother handover with high vehicle mobility.
	
	{{By combing these 5G technologies, SCSs can bring three-fold benefits of flexible deployment, green operation and enhanced QoS, paving the way to cost-effective vehicular networking.}}
	
	\begin{figure*}[t]
		\centering
		\includegraphics[width=4.5in]{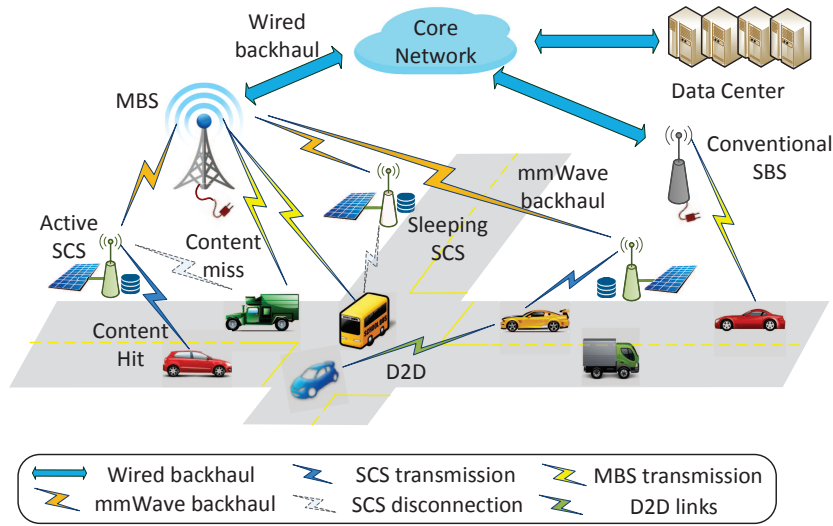}
		\caption{Vehicular network architecture with self-sustaining caching stations.}
		\label{fig_scenario}
	\end{figure*}
	
	\subsection{5G-Enabled Heterogeneous Vehicular Network Architecture}
	
	{{With SCSs integrated, a heterogeneous vehicular network architecture is formed and shown as Fig.~\ref{fig_scenario}.}}
	The conventional macro base stations (MBSs) and small cell base stations (SBSs) are connected with high speed wired backhauls and powered by conventional power grid, which mainly provide network coverage and control for reliability.
	Meanwhile, the SCSs are densely deployed for capacity enhancement, and mainly provide high speed data access based on stored contents.
	Furthermore, V2V communications are also enabled through device-to-device (D2D) links.
	{{The control and user planes are separated (i.e., C/U plane separation) for reliable and flexible access. Specifically, vehicles maintain dual connectivities: one with MBSs for signaling and control information, and the other with SCSs for high data rate transmissions or with other vehicles for instant message exchange. 
			As MBSs can provide ubiquitous signaling coverage with large cell radius, such a separation architecture can better support vehicle mobility with less frequent handover.}}
	
	The proposed architecture can support both safety and non-safety-related vehicular applications.
	For the safety-related use case, critical-event (such as collision or emerging stop) wanning messages can be exchanged locally via V2V communications with extremely low latency.
	For the non-safety-related use cases, MBSs and SCSs can enable better driving experiences, through services such as road condition broadcast, parking assistant and in-car infotainment. 
	In fact, the non-safety-related applications can be data hungry, and accounts for more than 90\% on-vehicle traffic \cite{Malandrino16_OnRoadFog_data}.
	Accordingly, we mainly focus on V2I (vehicle-to-MBS and vehicle-to-SCS) communications, and investigate cost-effective accommodation solutions for the increasing vehicular traffic demand.
	
	With SCSs offloading traffic from MBSs, the service process for vehicle users is as follows.
	The vehicle user can be directly served if its required content is stored at the associated SCS, which is called the \emph{content hit} case as shown in Fig.~\ref{fig_scenario}.
	Otherwise, the vehicle user is served by the MBSs, which is called the \emph{content miss} case.
	To serve the content miss users, the associated MBS needs to fetch contents from remote data centers via wired backhauls, according to the conventional cellular communication technologies.
	
	With sufficient cache size and well-designed caching schemes, deploying SCSs can effectively reduce the traffic load of MBSs.
	Furthermore, the content hit vehicle users can enjoy better quality of experience (QoE) with lower end-to-end delay.
	As such, the proposed network architecture can provide high capacity for vehicular communications at lower cost.
	
\section{Hierarchical Network Management Framework}
    \label{sec_management}
	\subsection{Management Challenges}
	
	
	\subsubsection{Network Heterogeneity}
	MBSs and SCSs exhibit distinct features with respect to coverage, user capacity, content access, etc.
	MBSs guarantee ubiquitous coverage with a large cell radius (e.g., several kilometers), and hence the associated users can enjoy less handover when moving at a high speed.
	However, the large coverage radius may also bring massive connections to each MBS.
	As a result, MBSs can only provide limited radio resources to each vehicle user at low transmission rates.
	On the contrary, each SCS covers relatively smaller area and serves less vehicle users at high transmission rates.
	Besides, the SCS users can get files without backhaul transmissions, which further reduces end-to-end delay.
	Nevertheless, SCSs mainly target on popular file transmission, and their small coverage radius may cause frequent handover issues.
	In addition to the heterogeneity of network infrastructures, vehicular services are at a wide-range requiring heterogeneous QoS requirements.
	For example, the safety and control messages are delay-sensitive but occupy limited radio resources, whereas non-safety-related applications such as social network on the wheel and map downloading requires large bandwidth but can endure longer delay. 
	The heterogeneity of network resource and traffic demand need be taken into consideration for user association and resource management.
	
	\subsubsection{Highly Dynamic Traffic Demand}
	
	On-road wireless traffic are highly dynamic in both time and spatial domains due to the variations of vehicle intensity.
	For example, the traffic volume during rush hours can be 90 times of that in late night, while the traffic volume on one direction can be 4 time of the opposite direction at the same road segment \cite{Bai09_vehicle_traffic_spatio_temporal}.
	Such traffic non-uniformity can pose great challenges to network management.
	Traffic bursty during rush hours may lead to service outage due to limited network resources, whereas network resources cannot be fully utilized during off-peak period.
	Besides, the spatial traffic imbalance may also lead to congestion in some cells while resource underutilized in other cells, degrading both service quality and network efficiency.
	In spite of traffic volume variation, the popularity distribution of different contents also varies with time, and thus SCSs need to update their content cache to maintain high content hit rate.
	
	
	\subsubsection{Intermittent Energy Supply}
	
	Unlike conventional power grid, renewable energy arrives randomly in an intermittent manner, which is likely to mismatch with traffic demand.
	For example, the solar powered SCSs cannot provide service after sunset well, when the vehicular networks may be still heavily loaded.
	On the contrary, on-road traffic can be very light at noon, while solar energy can be harvested at peak rate.
	The unbalanced power demand and supply can cause energy outage as well as battery overflow, which degrades system reliability and also leads to energy waste.
	Accordingly, energy sustainability is critical to the proposed network paradigm, which requires intelligent network management to minimize the probability of energy outage and overflow.
	
	{{In addition to the above mentioned challenges, there are also other issues need to be addressed, such as vehicle mobility, and time-varying mmWave backhaul capacity.
			To summarize, the network should fully utilize heterogeneous network resources to provide reliable on-demand service, so as to minimize operational cost while meeting differentiate QoS requirements of on-road mobile applications.}}

	\begin{figure*}[!t]
		\centering
		\includegraphics[width=6in]{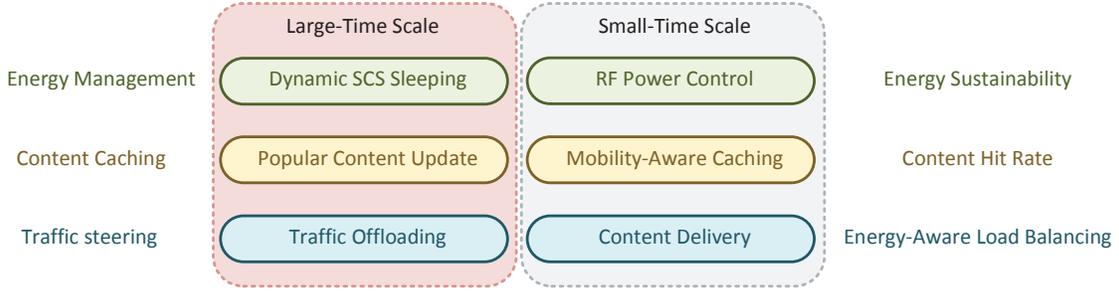}
		\caption{Hierarchical vehicular network management framework.}
		\label{fig_ManageFramework}
	\end{figure*}
	
	\subsection{Hierarchical Network Management}
	
	To address the above mentioned challenges, we propose a hierarchical network management framework, as shown in Fig.~\ref{fig_ManageFramework}.
	The proposed framework mainly includes three components: energy management, content caching, and traffic steering.
	Furthermore, network management are conducted in both large (e.g., minutes or hours) and small time scales (e.g., seconds) with different strategies.
	
	\emph{\textbf{Energy Management}} mainly deals with the randomness of renewable energy supply.
	Specifically, we propose dynamic SCS sleeping and Radio Frequency (RF) power control to reshape renewable energy supply by manipulating the process of charging and discharging.
	Notice that the power consumption of an SCS consists of two parts: (1) constant power which is irrelevant with traffic load, and (2) RF power which scales with traffic demand, through adjusting the transmit power level or the number of utilized subcarriers.
	RF power control can reduce the RF power consumed by wireless transmission, while dynamic SCS sleeping can further reduce the constant part by completely deactivating the SCS.
	Although dynamic SCS sleeping is more effective for power saving, frequent switching may cause additional cost.
	Thus, SCS sleeping can be performed in large time scale, and then each active SCS further adjusts the RF power in small time scale.
	Hierarchical energy management can reshape renewable energy supply in time domain to match the power demand at SCSs.
	For example, the SCSs with insufficient energy can switch to sleep mode, while active SCSs with oversupplied energy can enlarge transmit power to offload more vehicular traffic.
	In this way, SCSs can achieve energy-sustainable operation with balanced power demand and supply.
	
	{{\emph{\textbf{Content Caching}} schemes are critical for system performance, due to the limited storage capacity and constrained mmWave backhauls.
			Specifically, we consider two design objectives, i.e., content hit rate and mobility support.}}
	{{Content hit rate determines the maximal amount of traffic offloaded from MSBSs to SCSs, which reflects the service capability of SCSs.
			Meanwhile, mobility-aware caching can be implemented to realize seamless handover, where contents can be pro-actively fetched and stored at candidate cells based on handover prediction \cite{Wang16_cache_mobility_mag}.}}
	To realize these two objectives, the cache can be divided into two parts: one for the popular contents to guarantee content hit rate\footnote{Storing most popular contents can maximize content hit rate if SCSs do not cooperate with each other \cite{Gong16_push_cache_EH_ICC}.}, and the other for mobility-aware caching.
	Notice that mobility-aware caching requires frequent content fetching at the same time scale of vehicle handover, whereas the content popularity distribution may vary at a relatively slow pace.
	{{As the capacity of mmWave backhaul is constrained and varies dynamically with channel conditions, mobility-aware caching can be conducted timely in small time scale, whereas the popular contents can be updated in large time scale opportunistically based on channel status. Furthermore, each RSUs should update content based on their own locations, since the on-road mobile traffic requests can show location-based popularity.}}
	
	\emph{\textbf{Traffic Steering}} further reshapes traffic distribution to match the renewable energy supply, i.e., energy-aware load balancing.
	To this end, traffic offloading and content delivery are performed in different time scales, corresponding to energy management operations.
	In the large time scale, traffic offloading optimizes the amount of traffic served by each active SCS based on their renewable energy supply.
	For instance, SCSs with lower battery can serve fewer vehicle users, and vice versa.
	In the small time scale, energy-aware content delivery optimizes the transmission scheduling based on the SCS transmit power, to further improve QoS performance.
	For example, the delivery of best effort contents can be delayed when the transmit power is reduced, while SCSs can pro-actively push popular contents to vehicle users before requests when renewable energy is oversupplied.
	In essence, traffic offloading tunes the traffic load of each SCS (i.e., spatial traffic reshaping) while content delivery further adjusts traffic load at each time slot (i.e., temporal traffic reshaping).
	As such, traffic demand can be balanced with respect to renewable energy supply status.

	\begin{figure*}[!t]
		\centering
		\includegraphics[width=4in]{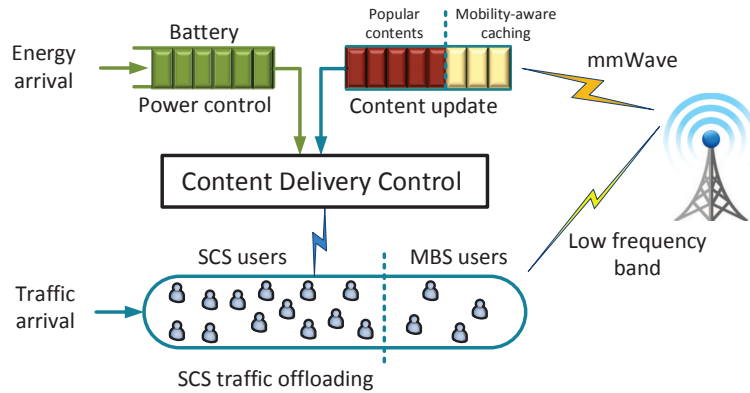}
		\caption{SCS management.}
		\label{fig_SCS}
	\end{figure*}
	
	Notice that these three operations jointly affects the performance of SCSs.
	For each SCS, content delivery control should be conducted based on the available battery, stored contents, and offloaded traffic status, as shown in Fig.~\ref{fig_SCS}.
	Therefore, the joint optimization of caching, energy and traffic management can help to improve system performance, at the price of higher operational complexity.
	
\section{Case Studies}
    \label{sec_CaseStudy}

	Under the proposed management framework, many implementation problems still need to be addressed, such as caching design, intelligent energy and traffic management.
	In this part, we introduce two specific design examples on caching size optimization and sustainable traffic-energy management, respectively.
	Numerical results will be presented to offer insights into practical network deployment and operations.
	
	We consider a two-way highway scenario where SCSs are deployed regularly with coverage radius of 500 m.
	The file library consists of 1000 files whose popularity distribution follows Zipf function with exponent $\gamma_\mathrm{f}$.
	The headway among neighboring vehicles follows exponential distribution of parameter $\lambda_\mathrm{v}$.
	In fact, $\lambda_\mathrm{v}$ reflects the vehicle density, and a larger $\lambda_\mathrm{v}$ characterizes denser vehicle scenarios.
	Assume all vehicles are greedy sources with average data rate requirement of 10 Mbps, and each SCS can simultaneously serve 10 vehicle users at most due to radio resource limitations.

	\subsection{Cache Size Design}
	
	\begin{figure}[!t]
		\centering
		\includegraphics[width=2.5in]{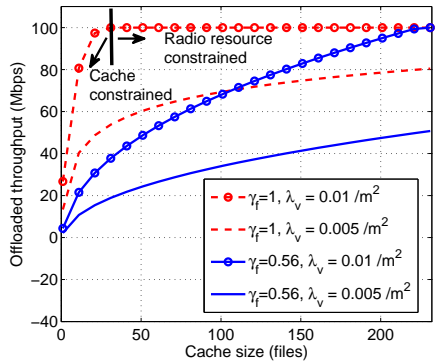}
		\caption{Service capability of SCSs with different cache sizes.}
		\label{fig_cache_size_influence}
	\end{figure}

	
	Fig.~\ref{fig_cache_size_influence} illustrates the amount of traffic that can be offloaded to each SCS under different traffic density $\lambda_\mathrm{v}$ and content popularity distributions $\gamma_\mathrm{f}$\footnote{$\gamma_\mathrm{f}=0.56$ comes from real data measurement of Youtube video streaming \cite{Gill07_youtube}, $\gamma_\mathrm{f}=1$ can describe location-based services (e.g., map downloading) whose requests may present higher similarity.}, which firstly increase but then level off with the increase of cache size.
	The reason is that the amount of offloaded traffic is also constrained by available radio resources\footnote{Notice that each SCS can simultaneously serve 10 users at most, each with data rate of 10 Mbps.}.
	Accordingly, the system performance can be divided into two regions, i.e., cache constrained region and radio resource constrained region, as shown in Fig.~\ref{fig_cache_size_influence}.
	In the cache constrained region, content hit rate is low and fewer vehicle users can be offloaded to SCSs, which corresponds to non-saturated case with under-utilized radio resources.
	As the cache size increases, more users can be offloaded to SCSs with higher content hit rate.
	Accordingly, the traffic of SCSs become saturated, and the throughput of SCS no longer increases due to the limitation of available radio resource.
	
	The obtained results reveal the Pareto optimality of cache size and SCS density, and offer insights into practical network deployment.
	For example, the optimal cache size should be larger than 31 files when the SCS coverage is 500 m, vehicle density is 0.01 /m, and popularity parameter $\gamma_\mathrm{f} = 1$.
	Furthermore, the cost-optimal combination of cache size and SCS density for the given network capacity can be also found, given the cost functions of cache size and SCSs.

	\subsection{Sustainable Traffic-Energy Management}
	
	To reveal the importance of sustainable traffic-energy management, we study the service capability of the SCS under different traffic-energy management schemes.
	{{The greedy scheme is adopted as a baseline, where the SCSs always keep active and work at the maximal transmit power.}}
	With the sustainable traffic-energy management, an SCS goes into sleep if the available energy is insufficient to support its constant power consumption, {{otherwise it stays active and adjusts the transmit power and offloaded traffic amount based on the instant energy arrival rate.}}
	The redundant energy is saved in battery for future use, and the battery capacity is considered to be large enough without overflow.
	
\begin{figure*}[!t]
		\centering
		\subfloat[Daily traffic and renewable energy profiles] {\includegraphics[width=2.5in]{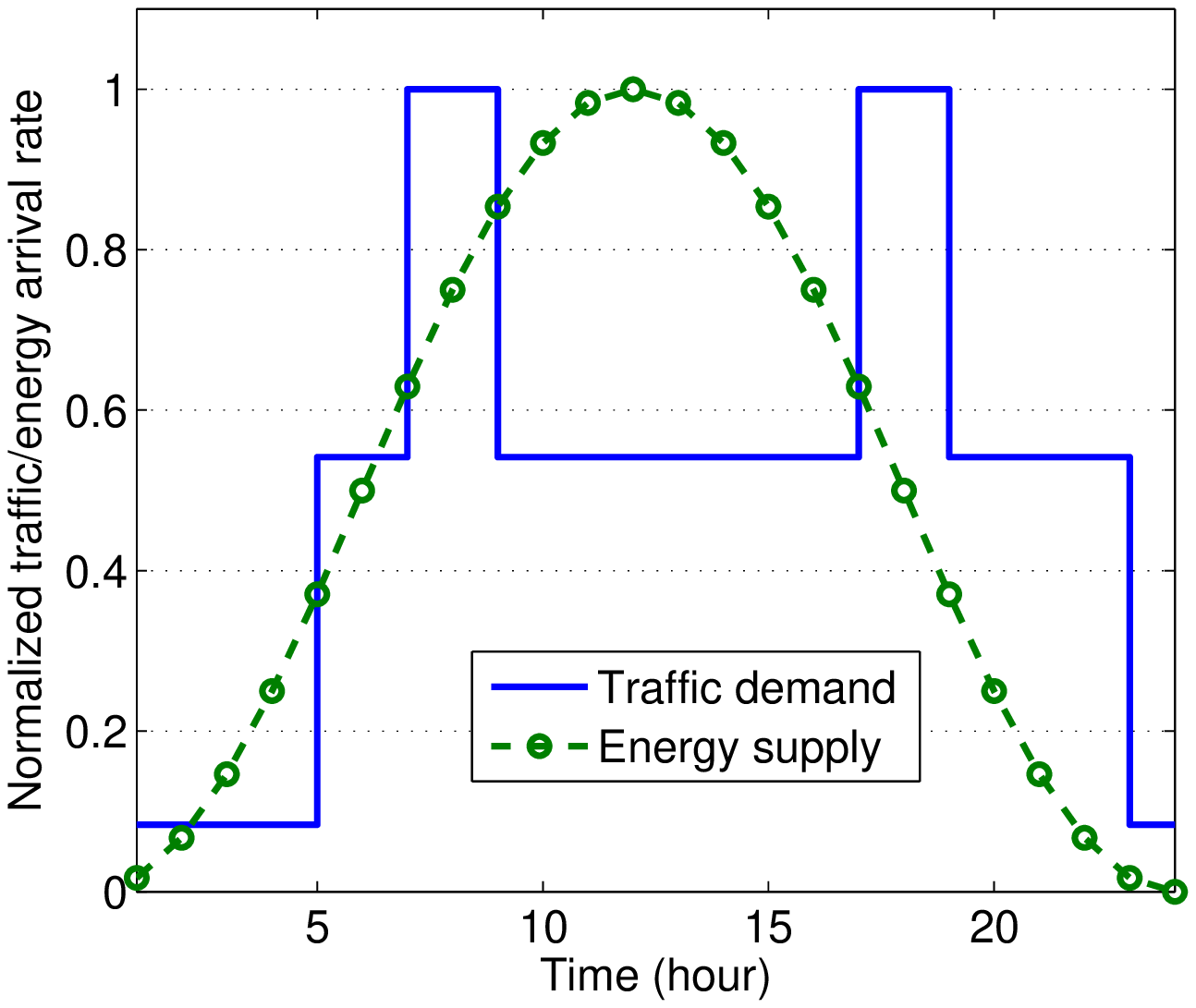}
			\label{fig_traffic_energy}}
		\hfil
		\subfloat[Normalized offloaded traffic]{\includegraphics[width=2.5in]{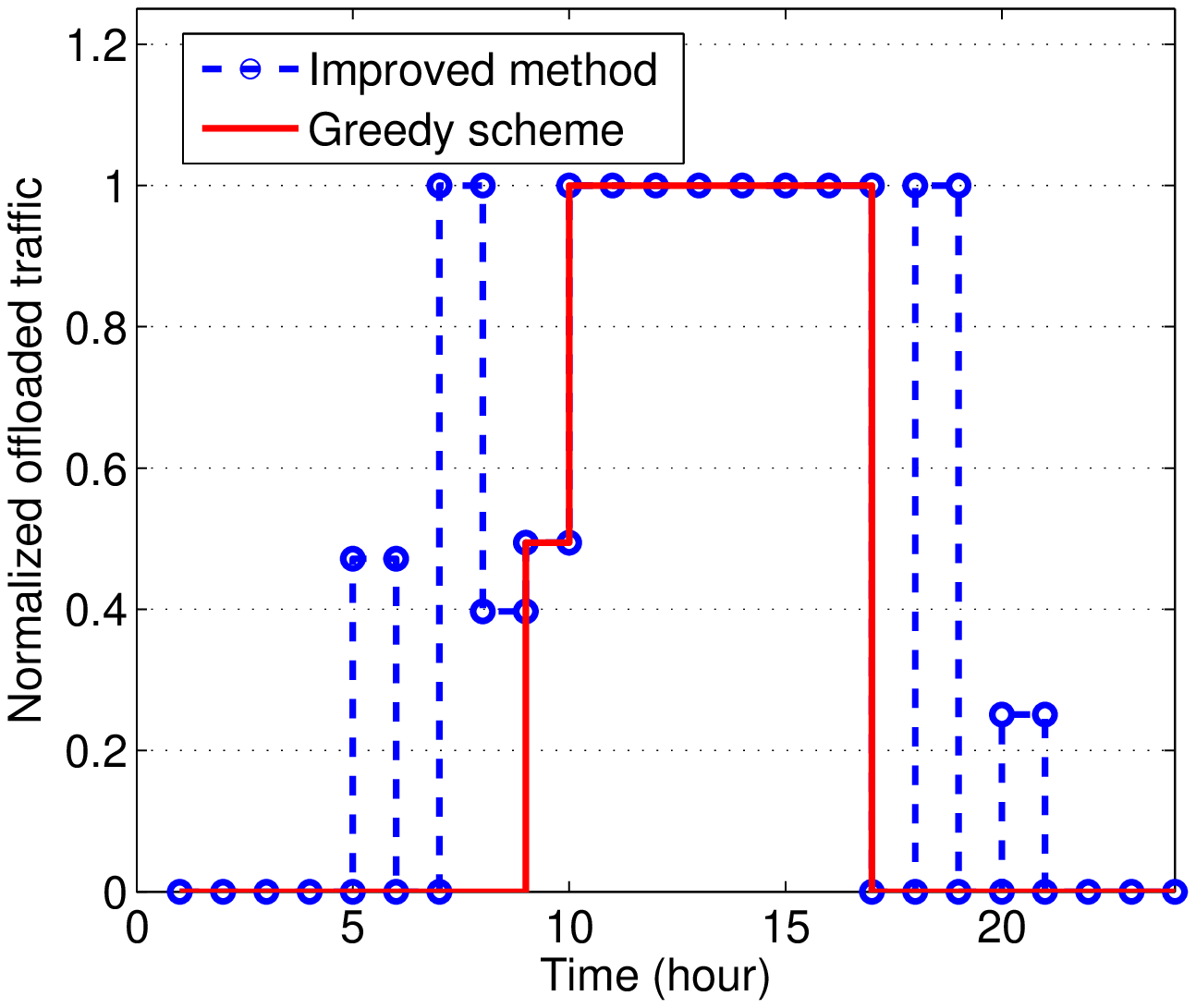}
			\label{fig_performance_comparison}}
		\caption{System performance with sustainable traffic-energy management.}
		\label{fig_evaluation}
	\end{figure*}

	Fig.~\ref{fig_evaluation}(a) illustrates the normalized traffic and energy profiles.
	Specifically, the two peaks of the traffic profile correspond to the on-road rush hours in the morning and afternoon respectively.
	Meanwhile, solar energy harvesting is considered, and the daily energy arrival rate is modeled as a sine function with a peak at noon.
	Under the considered traffic and energy profiles, the normalized offloaded traffic (i.e., the percentage of vehicle users offloaded to the SCSs) is shown as Fig.~\ref{fig_evaluation}(b), where the peak energy arrival rate equals to the maximal power consumption of each SCS and the highest traffic density corresponds to the SCS capacity.
	As shown in Fig.~\ref{fig_evaluation}(b), the sustainable traffic-energy management method outperforms the greedy scheme.
	Specifically, the sustainable traffic-energy management can increase SCS capacity to nearly 1.7 times compared with the greedy scheme, realizing cost-effective management.
	{{In fact, the greedy scheme can minimize the probability of battery overflow, which performs well with sufficient energy supply. The sustainable traffic-energy management scheme further reduces the probability of battery outage through dynamic SCS sleeping, which can better utilize energy with higher efficiency.}}
	
\section{Open Research Issues}
    \label{sec_research_topic}

	    As the study on cost-effective vehicular network is still at the infant age, there are many research issues remaining unsolved. 
	    
	    \subsubsection{Caching Scheme Design}
	    
	    Under the proposed management framework, efficient caching schemes should be designed to maximize content hit rate while minimizing handover cost, by determining caching size splitting, popular content update, and mobility-aware caching.
	    For the given caching splitting, the problems of popular content update and mobility-based caching can be both modeled as a Markov Decision Process (MDP), and dynamic programing or machine learning provide powerful solutions.
	    Then, the optimal cache splitting can be further explored based on the designed schemes of popular content update and mobility-aware caching schemes. 
	    Notice that there exists a tradeoff between content hit rate and handover delay with different caching size splitting ratios.
	    Accordingly, Pareto optimality can serve as the design criteria.
	    
	    \subsubsection{Sustainable Traffic and Energy Management}
	    
	    As demonstrated in the case study, the conventional greedy traffic offloading and energy management schemes are insufficient, due to the randomness of renewable energy and highly dynamic vehicular traffic.
	    {{Sustainable traffic and energy management is desired to balance power demand and supply at each SCS, through the cooperation among neighboring SCSs and cellular networks.}}
	    An optimization problem can be formulated to maximize the service capability SCSs, subject to energy casualty and QoS requirements of all users.
	    The decision variables include the work mode, offloaded traffic amount, transmit power, and content delivery scheduling of each SCS.
	    However, this problem can be extremely complex due to the multi-dimensional coupled optimization variables.
	    In this case, the hierarchical management framework can be exploited for problem decoupling.
	    Specifically, we can deal with the work mode and offloaded traffic amount in the large time scale, while adjust the transmit power and schedule content delivery in the small time scale.
	    Then, low-complexity management schemes can be proposed for practical implementations.
	    
	    \subsubsection{Cost-Effective SCS Deployment}
	    
	    The introduction of SCSs also poses new design issues for network deployment, as discussed in the case study of cache size optimization.
	    In fact, the service capability of SCSs can increase with denser SCS, larger cache size, or higher battery capacity.
	    Accordingly, cost-effective SCS deployment should jointly optimize these system parameters to minimize the long-term network cost while meeting vehicular traffic demand.
	    Specifically, the tradeoff among those system parameters should be carefully studied to obtain the cost-optimal combination.
	    Stochastic geometry can be adopted for such large-scale system performance analysis, which can provide favorable closed-form results with reasonable approximations \cite{mine_TWC_SCoff}.

    
\section{Conclusions}
	\label{sec_conclusions}

	We have introduced a new type of on-road base station, namely SCS, to exploit renewable energy harvesting, mmWave backhaul, and content caching techniques to achieve flexible, sustainable, and cost-effective vehicular networking.
	With promising 5G technologies, SCSs can enable ``drop-and-play'' deployment, green operation, and low-latency content delivery, paving the way to cost-effective vehicular networking.
	Furthermore, a heterogeneous vehicular network architecture has been proposed to provide high capacity and better QoS to vehicle users, by efficiently exploring the specific advantages of SCSs and MBSs.
	In addition, a hierarchical management framework has been designed, where energy management, content caching, and traffic steering are performed in both large and small time scales to deal with the dynamics in energy supply and traffic demand.
	Case studies on cache size optimization and sustainable traffic-energy management have been conducted to provide insights into practical design of 5G-enabled vehicular networks. 
	Moreover, important research topics on SCSs have also been discussed.


\bibliographystyle{IEEEtran}


\end{document}